\documentclass[preprintnumbers,amsmath,amssymb,showpacs,twocolumn]{revtex4}

\usepackage{graphicx}
\usepackage{dcolumn}
\usepackage{bm}

\begin{document}

\title{Generation of Six-Qubit Cluster State in Ion-Trap System}

\author{Li Shu-Yue, Yan Feng-Li}
\thanks{flyan@mail.hebtu.edu.cn\\Supported by the National Natural Science Foundation of China under
Grant No: 10971247, Hebei Natural Science Foundation of China under
Grant No: F2009000311 and  the Key Project of Science and Technology
Research of Education Ministry of China under Grant No: 207011.}
            \affiliation { College of Physics Science and Information
            Engineering, Hebei Normal University, Shijiazhuang 050016, China\\
Hebei Advanced Thin Films Laboratory,  Shijiazhuang 050016, China}

\date{December 23, 2009}

\begin{abstract}
Based on resonant sideband excitation, we present a scheme for the
generation of six-qubit cluster state in ion-trap system. One can
realize experimentally this scheme  with presently available
techniques.
\end{abstract}

\pacs{42.50.Dv}

\maketitle

\section {Introduction}

Recently Briegel et al.  introduced cluster states   as a
fundamental resource aimed at the linear optics one way quantum
computation \cite{BriegelRaussendorf, RaussendorfBriegel}. Walther
et al. \cite {WaltherReschRudolph} and Tame et al. \cite
{TamePrevedelPaternostro}   demonstrated independently the
experimental feasibility of one way computation by four-photon
cluster states.

Many schemes for the generation of  four-qubit cluster states have
been proposed so far \cite {ZhengSB, VallonePomaricoMataloni,
ZhangGaoFeng, BarrettKok}. In this paper, based on resonant sideband
excitation   we generalize  Zheng's scheme for the generation of
four-qubit cluster state \cite {ZhengSB} to the case of six-qubit
to present a scheme for the generation of six-qubit cluster state
in ion-trap system. It can be realized with presently available
experimental techniques.

\section {Generation of Six-Qubit Cluster State in Ion-Trap System}

The $N$-qubit cluster states can be written in the form
\begin{equation}
|\Psi_N\rangle=\frac
{1}{2^{N/2}}\otimes^{N}_{\alpha=1}(|0\rangle_\alpha\sigma_{z}^{(\alpha+1)}+|1\rangle_\alpha),
\end{equation}
with the convention $\sigma_z^{(N+1)}\equiv 1$. Here $\sigma_z=|0\rangle\langle0|-|1\rangle\langle1|$ is
Pauli operator.

Suppose  that  six ions are confined in the ion-trap system and each
ion has two excited metastable states $|e\rangle$ and $|e'\rangle$
and one ground state $|g\rangle$. The state of the qubit is
$\alpha|e\rangle+\beta|g\rangle$, where $\alpha$ and $\beta$ are
complex numbers satisfying $|\alpha|^2+|\beta|^2=1$. We prepare
initially the first ion  in the state $|e_1\rangle$ and the
center-of-mass vibrational mode  in the vacuum state $|0\rangle$.
When a laser tuned to the first lower vibrational sideband was
applied to the first ion, it induces  the transition between
$|g_1\rangle|1\rangle$ and $|e_1\rangle|0\rangle$. Assume the laser
is off resonant with the transition $|g_1\rangle|1\rangle$
$\rightarrow$ $|e'_1\rangle|0\rangle$ and hence this laser will
leave the state $|e'_1\rangle$ unaffected during the interaction.
When the Lamb-Dicke criterion is satisfied, i.e. the Lamb-Dicke
parameter $\eta\ll 1$, and in the weak-excitation regime, where Rabi
frequency $\Omega$ is much smaller than vibrational frequency, the
Hamiltonian in an interaction picture reads
\cite{CiracZoller,ZhengSB}
\begin{equation}
H=\mathrm{i}\frac {\eta}{2}\Omega {\rm e}^{-{\rm
i}\phi}a|e_1\rangle\langle g_1|+\texttt{h.c}.,
\end{equation}
where $a^+$ and $a$ are the creation and annihilation operators of
the center-of-mass vibrational mode of the trapped ions,   $\phi$ is
the  phase of this laser field. As the Hamiltonian for the quantum
system is not a time-varying,  hence if the laser beam is on for a
certain time $t$, the evolution of the system will be described by
the unitary operator
\begin{equation}
\begin{array}{ll}
U(t)&={\rm e}^{-\mathrm{i}Ht}\\
&=\exp\{-\texttt{i(i}\frac {\eta}{2}\Omega {\rm e}^{-{\rm
i}\phi}a|e_1\rangle\langle g_1|+\texttt{h.c.})t\}\\
&= \cos(\frac {\eta\Omega t}{2}\sqrt {1+a^+a})|e_1\rangle\langle e_1|\\
&~~+ \mathrm{e}^{-\mathrm{i}\phi}\frac {\sin(\frac {\eta\Omega
t}{2}\sqrt {1+a^+a})}{\sqrt {1+a^+a}}a|e_1\rangle\langle g_1|\\
&~~-\mathrm{e}^{\mathrm{i}\phi}\frac {\sin(\frac {\eta\Omega
t}{2}\sqrt {a^+a})}{\sqrt {a^+a}}a^+ |g_1\rangle\langle e_1|\\
&~~+ \cos(\frac{\eta\Omega t}{2}\sqrt {a^+a})|g_1\rangle\langle
g_1|.\end{array}
\end{equation}
 After an interaction time $\tau_1$, the state of the system consisted of this ion and the center-of-mass
mode evolves  into
\begin{equation}
U(\tau_1)|e_1\rangle|0\rangle=\cos(\frac {\eta}{2}\Omega
\tau_1)|e_1\rangle|0\rangle-\mathrm{e}^{\mathrm{i} \phi}\sin(\frac
{\eta}{2}\Omega \tau_1)|g_1\rangle|1\rangle.
\end{equation}
By choosing  $\phi=\pi$ and $\eta\Omega \tau_1=\pi/2$,  one has
\cite {ZhengSB}
\begin{equation}
\frac {1}{\sqrt 2}(|e_1\rangle|0\rangle+|g_1\rangle|1\rangle).
\end{equation}

 Following the ideas introduced in Ref.\cite {ZhengSB,CiracZoller}, we now drive the third ion
  with a laser tuned to the first lower
 vibrational sideband with respect to the transition $|g_3\rangle|1\rangle$
 $\rightarrow$ $|e'_3\rangle|0\rangle$. In this case, the interaction
 Hamiltonian in an interaction picture is
  \begin{equation}
H'=\mathrm{i}\frac {\eta}{2}\Omega' \mathrm{e}^{-\mathrm{i}
\phi'}a|e'_3\rangle\langle g_3|+\texttt{h.c.},
 \end{equation}
where parameters $\Omega'$,  $\phi'$ have the similar meaning with
$\Omega$, $\phi$ respectively.

If the laser beam is on for a certain time $t$, the evolution
operator of the system  combined by the third ion and the
center-of-mass mode reads
\begin{equation}
\begin{array}{ll}
U'(t)&={\rm e}^{-\mathrm{i}H't}\\
&=\exp\{-\texttt{i(i}\frac {\eta}{2}\Omega' {\rm e}^{-{\rm
i}\phi'}a|e'_3\rangle\langle g_3|+\texttt{h.c.})t\}\\
&= \cos(\frac {\eta\Omega' t}{2}\sqrt {1+a^+a})|e'_3\rangle\langle e'_3|\\
&~~+ \mathrm{e}^{-\mathrm{i}\phi'}\frac {\sin(\frac {\eta\Omega'
t}{2}\sqrt {1+a^+a})}{\sqrt {1+a^+a}}a|e'_3\rangle\langle g_3|\\
&~~-\mathrm{e}^{\mathrm{i}\phi'}\frac {\sin(\frac {\eta\Omega'
t}{2}\sqrt {a^+a})}{\sqrt {a^+a}}a^+ |g_3\rangle\langle e'_3|\\
&~~+ \cos(\frac{\eta\Omega' t}{2}\sqrt {a^+a})|g_3\rangle\langle
g_3|.\end{array}
\end{equation}

After an interaction time $\tau_2$, the state $|g_3\rangle|1\rangle$
evolves to
\begin{equation}
U'(\tau_2)|g_3\rangle |1\rangle=\cos(\frac {\eta}{2}\Omega'
\tau_2)|g_3\rangle|1\rangle+\mathrm{e}^{-\mathrm{i} \phi'}\sin(\frac
{\eta}{2}\Omega' \tau_2)|e'_3\rangle|0\rangle.
\end{equation}
By choosing $\eta\Omega'\tau_2 =2\pi$,  we have
 $|1\rangle|g_3\rangle \rightarrow -|1\rangle|g_3\rangle$,
 but other states do not unaffect during the interaction.
  This is just a phase gate between the third ion and the center-of-mass vibrational mode,  proposed
   by Cirac and Zoller \cite {ZhengSB, CiracZoller}.

   If the third ion is prepared  initially  in the state $\frac {1}{\sqrt
 2}(|g_3\rangle-|e_3\rangle)$, the system consisting of the first ion and the center-of-mass mode is in the state
 given by  Eq.(5), then after the phase gate operation between the third ion and the center-of-mass
 mode has been implemented, the state of the system combined by the first ion, the third ion and the
 center-of-mass mode is in  \cite {ZhengSB}
   \begin{equation}
\frac
{1}{2}[|e_1\rangle(|g_3\rangle-|e_3\rangle)|0\rangle-|g_1\rangle(|g_3\rangle+|e_3\rangle)|1\rangle].
   \end{equation}

Suppose the second ion is prepared initially in the state
$|g_2\rangle$. Now one drives the second ion with a laser tuned to
the first lower vibrational sideband with respect to the transformation
$|g_2\rangle|1\rangle \rightarrow |e_2\rangle|0\rangle$, $|g_2\rangle|0\rangle \rightarrow |g_2\rangle|0\rangle$.
 After that  the state of the system consisting of the first ion, the second
ion, the third ion and the center-of-mass mode becomes \cite
{ZhengSB}
\begin{equation}
\frac
{1}{2}[|e_1\rangle|g_2\rangle(|g_3\rangle-|e_3\rangle)-|g_1\rangle|e_2\rangle(|e_3\rangle+|g_3\rangle)]|0\rangle.
\end{equation}

It is easy to obtain the state \cite {ZhengSB}
\begin{equation}
\begin{array}{l}
\frac {1}{2\sqrt
2}[(|g_1\rangle+|e_1\rangle)|g_2\rangle(|g_3\rangle-|e_3\rangle)\\
-(|g_1\rangle-|e_1\rangle)|e_2\rangle(|e_3\rangle+|g_3\rangle)]|0\rangle.
\end{array}\end{equation} by making the following unitary
operation
\begin{equation}
|e_1\rangle \rightarrow \frac {1}{\sqrt 2}(|e_1\rangle+|g_1\rangle),
|g_1\rangle \rightarrow \frac {1}{\sqrt 2}(|g_1\rangle-|e_1\rangle).
\end{equation}

Assume that the fourth ion is prepared initially in the state $\frac
{1}{\sqrt 2}(|g_4\rangle-|e_4\rangle)$. One maps  this state to the
vibrational mode by applying a laser tuned to the first lower
vibrational sideband  to the fourth ion, that induces  the
transformation  $|e_4\rangle|0\rangle\rightarrow |g_4\rangle|1\rangle$, $|g_4\rangle|0\rangle\rightarrow |g_4\rangle|0\rangle$, one obtains the state
\begin{equation}
\begin{array}{l}\frac
{1}{4}[(|g_1\rangle+|e_1\rangle)|g_2\rangle(|g_3\rangle-|e_3\rangle)\\
-(|g_1\rangle-|e_1\rangle)|e_2\rangle(|e_3\rangle+|g_3\rangle)]|g_4\rangle(|0\rangle-|1\rangle).
\end{array}\end{equation}

By completing the phase gate operation between the third ion and the
vibrational mode,  the state given by Eq.(13) becomes
\begin{equation}
\begin{array}{l}
\frac
{1}{4}\{[(|g_1\rangle+|e_1\rangle)|g_2\rangle(|g_3\rangle-|e_3\rangle)\\
-(|g_1\rangle-|e_1\rangle)|e_2\rangle(|e_3\rangle+|g_3\rangle)]|0\rangle\\
-[(|g_1\rangle+|e_1\rangle)|g_2\rangle(-|g_3\rangle-|e_3\rangle)\\
-(|g_1\rangle-|e_1\rangle)|e_2\rangle(|e_3\rangle-|g_3\rangle)]|1\rangle\}|g_4\rangle.
\end{array}\end{equation}

Assume that the fifth ion is  initially in the state $\frac {1}{\sqrt
2}(|g_5\rangle-|e_5\rangle)$. Now we apply a laser tuned to the first lower
vibrational sideband, that induces  a phase gate between the fifth ion and the center-of-mass vibrational mode,  $|g_5\rangle|1\rangle\rightarrow -|g_5\rangle|1\rangle$. Then we
have
\begin{equation}
\begin{array}{l}
\frac {1}{4\sqrt
2}\{[(|g_1\rangle+|e_1\rangle)|g_2\rangle(|g_3\rangle-|e_3\rangle)\\
-(|g_1\rangle-|e_1\rangle)|e_2\rangle(|e_3\rangle+|g_3\rangle)](|g_5\rangle-|e_5\rangle)|0\rangle\\
+[(|g_1\rangle+|e_1\rangle)|g_2\rangle(-|g_3\rangle-|e_3\rangle)\\
-(|g_1\rangle-|e_1\rangle)|e_2\rangle(|e_3\rangle-|g_3\rangle)](|g_5\rangle+|e_5\rangle)|1\rangle\}|g_4\rangle.
\end{array}
\end{equation}

Now we map the state of the vibration mode to the fourth ion. By applying a laser tuned to the first lower vibrational
sideband to the fourth ion, we obtain the transformation $|g_4\rangle|1\rangle \rightarrow |e_4\rangle|0\rangle$, $|g_4\rangle|0\rangle \rightarrow |g_4\rangle|0\rangle$. Then we have
\begin{equation}
\begin{array}{l}\frac {1}{4\sqrt
2}\{[(|g_1\rangle+|e_1\rangle)|g_2\rangle(|g_3\rangle-|e_3\rangle)\\
-(|g_1\rangle-|e_1\rangle)|e_2\rangle(|e_3\rangle+|g_3\rangle)]|g_4\rangle(|g_5\rangle-|e_5\rangle)\\
+[(|g_1\rangle+|e_1\rangle)|g_2\rangle(-|g_3\rangle-|e_3\rangle)\\
-(|g_1\rangle-|e_1\rangle)|e_2\rangle(|e_3\rangle-|g_3\rangle)]|e_4\rangle(|g_5\rangle+|e_5\rangle)\}|0\rangle.
\end{array}\end{equation}

Suppose that the sixth ion is initially in the state  $\frac {1}{\sqrt
2}(|e_6\rangle+|g_6\rangle)$. By applying a laser tuned to the first lower vibrational
sideband to  this ion, one can map the  state of the sixth ion to the vibrational
mode.  So one has the state
\begin{equation}
\begin{array}{l}\frac
{1}{8}\{[(|g_1\rangle+|e_1\rangle)|g_2\rangle(|g_3\rangle-|e_3\rangle)\\
-(|g_1\rangle-|e_1\rangle)|e_2\rangle(|e_3\rangle+|g_3\rangle)]|g_4\rangle(|g_5\rangle-|e_5\rangle)\\
+[(|g_1\rangle+|e_1\rangle)|g_2\rangle(-|g_3\rangle-|e_3\rangle)\\
-(|g_1\rangle-|e_1\rangle)|e_2\rangle(|e_3\rangle-|g_3\rangle)]|e_4\rangle(|g_5\rangle+|e_5\rangle)\}|g_6\rangle\\
(|0\rangle+|1\rangle).
\end{array}\end{equation}

We now perform a phase gate operation between the fifth ion and
the vibrational mode, leading to
\begin{equation}
\begin{array}{l}\frac
{1}{8}(\{[(|g_1\rangle+|e_1\rangle)|g_2\rangle(|g_3\rangle-|e_3\rangle)\\
-(|g_1\rangle-|e_1\rangle)|e_2\rangle(|e_3\rangle+|g_3\rangle)]|g_4\rangle(|g_5\rangle-|e_5\rangle)\\
+[(|g_1\rangle+|e_1\rangle)|g_2\rangle(-|g_3\rangle-|e_3\rangle)\\
-(|g_1\rangle-|e_1\rangle)|e_2\rangle(|e_3\rangle-|g_3\rangle)]|e_4\rangle(|g_5\rangle+|e_5\rangle)\}|0\rangle\\
+\{[(|g_1\rangle+|e_1\rangle)|g_2\rangle(|g_3\rangle-|e_3\rangle)\\
-(|g_1\rangle-|e_1\rangle)|e_2\rangle(|e_3\rangle+|g_3\rangle)]|g_4\rangle(-|g_5\rangle-|e_5\rangle)\\
+[(|g_1\rangle+|e_1\rangle)|g_2\rangle(-|g_3\rangle-|e_3\rangle)\\
-(|g_1\rangle-|e_1\rangle)|e_2\rangle(|e_3\rangle-|g_3\rangle)]|e_4\rangle(-|g_5\rangle+|e_5\rangle)\}|1\rangle).
\end{array}\end{equation}
Mapping the state of the vibrational mode  to  the sixth ion, we
have
\begin{equation}
\begin{array}{l}\frac
{1}{8}(\{[(|g_1\rangle+|e_1\rangle)|g_2\rangle(|g_3\rangle-|e_3\rangle)\\
-(|g_1\rangle-|e_1\rangle)|e_2\rangle(|e_3\rangle+|g_3\rangle)]|g_4\rangle(|g_5\rangle-|e_5\rangle)\\
+[(|g_1\rangle+|e_1\rangle)|g_2\rangle(-|g_3\rangle-|e_3\rangle)\\
-(|g_1\rangle-|e_1\rangle)|e_2\rangle(|e_3\rangle-|g_3\rangle)]|e_4\rangle(|g_5\rangle+|e_5\rangle)\}|g_6\rangle\\
+\{[(|g_1\rangle+|e_1\rangle)|g_2\rangle(|g_3\rangle-|e_3\rangle)\\
-(|g_1\rangle-|e_1\rangle)|e_2\rangle(|e_3\rangle+|g_3\rangle)]|g_4\rangle(-|g_5\rangle-|e_5\rangle)\\
+[(|g_1\rangle+|e_1\rangle)|g_2\rangle(-|g_3\rangle-|e_3\rangle)\\
-(|g_1\rangle-|e_1\rangle)|e_2\rangle(|e_3\rangle-|g_3\rangle)]|e_4\rangle(-|g_5\rangle+|e_5\rangle)\}|e_6\rangle)|0\rangle.
\end{array}\end{equation}
We can rewrite the state of the six ions stated in Eq.(19) as
\begin{equation}\begin{array}{l}
\frac
{1}{8}(|g_1\rangle\sigma_z^2+|e_1\rangle)(|g_2\rangle\sigma_z^3+|e_2\rangle)(|g_3\rangle\sigma_z^4+|e_3\rangle)\\
(|g_4\rangle\sigma_z^5+|e_4\rangle)(|g_5\rangle\sigma_z^6+|e_5\rangle)(|g_6\rangle+|e_6\rangle),
\end{array}\end{equation} where $\sigma_z^i=|g_i\rangle\langle
g_i|-|e_i\rangle\langle e_i|$. The state of Eq.(20) is just a
six-qubit cluster state.

Now we estimate the fidelity of our scheme. Obviously, our scheme  consists of
eight sideband excitations which couple the internal and external
degrees of freedom and trivial single-qubit operations. The fidelity
of each sideband excitation is about 0.93 \cite
{SchmidtKalerHaoffnerRiebe,RoosRiebeHaffner}, so the fidelity of our
whole procedure is about 0.56. We expect that the fidelity of each
sideband excitation can be improved. In that case, the fidelity of
our scheme can be enhanced.

\section {Summary}

We have proposed  a scheme for the generation of
six-qubit cluster state with trapped ions. The required
experimental techniques are presently available.
 We hope the scheme will be useful in quantum information proceeding.

\end{document}